\documentclass[prl,preprint,amssymb,amsfonts]{revtex4}
\usepackage{graphicx}
\usepackage{bm}

\def\beq{\begin{equation}}
\def\eeq{\end{equation}}
\begin{document}

\noindent
{\bf Comment on ``Ferroelectrically Induced Weak Ferromagnetism by Design''}\\

The question of how ferroelectric polarization is coupled to magnetism
in magnetoelectric multiferroics is of considerable current interest.
A recent Letter~\cite{fennie} analyzes the important ``ABO$_3$'' class
of perovskite multiferroics.  A symmetry argument is presented that
materials with antiferromagnetism on the B site lack
linear magnetoelectric coupling $E_{PLM}={\bf P} \cdot ({\bf L} \times
{\bf M})$ in the free energy, unlike A-site
antiferromagnets~\cite{fennie,edererfennie}.  Here ${\bf P}$ is
polarization and ${\bf L}$ and ${\bf M}$ are antiferromagnetic and
ferromagnetic moments.

This Comment presents a distinct analysis of $E_{PLM}$ in ABO$_3$
multiferroics.
We show that the argument of
Ref.~\onlinecite{fennie} does forbid $E_{PLM}$ if the final low-symmetry phase contains only one distortion that, like ${\bf P}$, breaks all inversion symmetries.
In reality there are multiple distortions in this symmetry class, and
cross-terms generate $E_{PLM}$ even in B-site materials, although the mechanism is different than in A-site materials.  Additional differences can emerge between A-site and B-site materials under dynamical assumptions beyond the static free energy.

For concreteness we use BiFeO$_3$, whose electronic structure has been
extensively studied~\cite{edererspaldin,ravindran}.  Below 1100K it
shows ferroelectric order with space group R3c and a 10-atom unit
cell.  In Ref.~\onlinecite{fennie}, the free energy is expanded from a
model R${\bar 3}$c phase with inversion center on Fe.
For now, assume that the polarization ${\bf P}$ arises from relative motion of the ionic
sublattices~\cite{edererspaldin} along the threefold axis $\hat{\bm{z}}$.
There is a 13.8$^\circ$ rotation $\alpha$ of the O octahedra
around Fe atoms ($\alpha {\bf \hat z}$ is ${\bf D}$ in
Ref.~\onlinecite{fennie}), which is even under Fe-site inversion $I_B$
and odd under Bi-site inversion $I_A$.  
However, distortions beyond ${\bf P}$ and $\alpha$ must
be present, either by counting R3c degrees of freedom or because
the 1.4$^\circ$ counter-rotation of upper and lower triangles within
an octahedron~\cite{kubel,note} cannot be obtained
by combining ${\bf P}$ and $\alpha$.

Coordinates for an oxygen distortion $\beta$ (Fig.~1) that 
causes this counter-rotation and is {\it orthogonal} to ${\bf P}$ are provided in EPAPS.
The $\beta$ distortion, like ${\bf P}$, is odd under $I_A$ and $I_B$.
For A-site magnetism, we have inversion
eigenvalues $I_A \bm{L}=+\bm{L}$, and $I_B \bm{L}=-\bm{L}$.  For
B-site magnetism, $I_A \bm{L}=-\bm{L}$ and $I_B \bm{L}=+\bm{L}$.  Allowed ${\bm L} \times {\bm M}$ terms are, with constants $\lambda_1, \lambda_2$,
\begin{eqnarray}
F_A &=& \lambda_1 \alpha \bm{P} \cdot (\bm{L}\times \bm{M}) + \lambda_2 \alpha\beta \hat{\bm{z}}\cdot (\bm{L}\times \bm{M}), \cr
F_B &=& \lambda_1 \alpha\beta \bm{P} \cdot (\bm{L}\times \bm{M})+
\lambda_2 \alpha \hat{\bm{z}} \cdot (\bm{L}\times \bm{M}).
\end{eqnarray}

More generally, ${\bf P}$ is some $I_A=I_B=-1$ distortion, and $E_{PLM}$ results from cross-terms between ${\bf P}$ and other such distortions, here $\beta$.  Magnetoelectric coupling should be sought in B-site structures with large $\beta$-type distortions as well as in A-site structures with large $\alpha$.



\begin{figure}
  \includegraphics[width=2.5in]{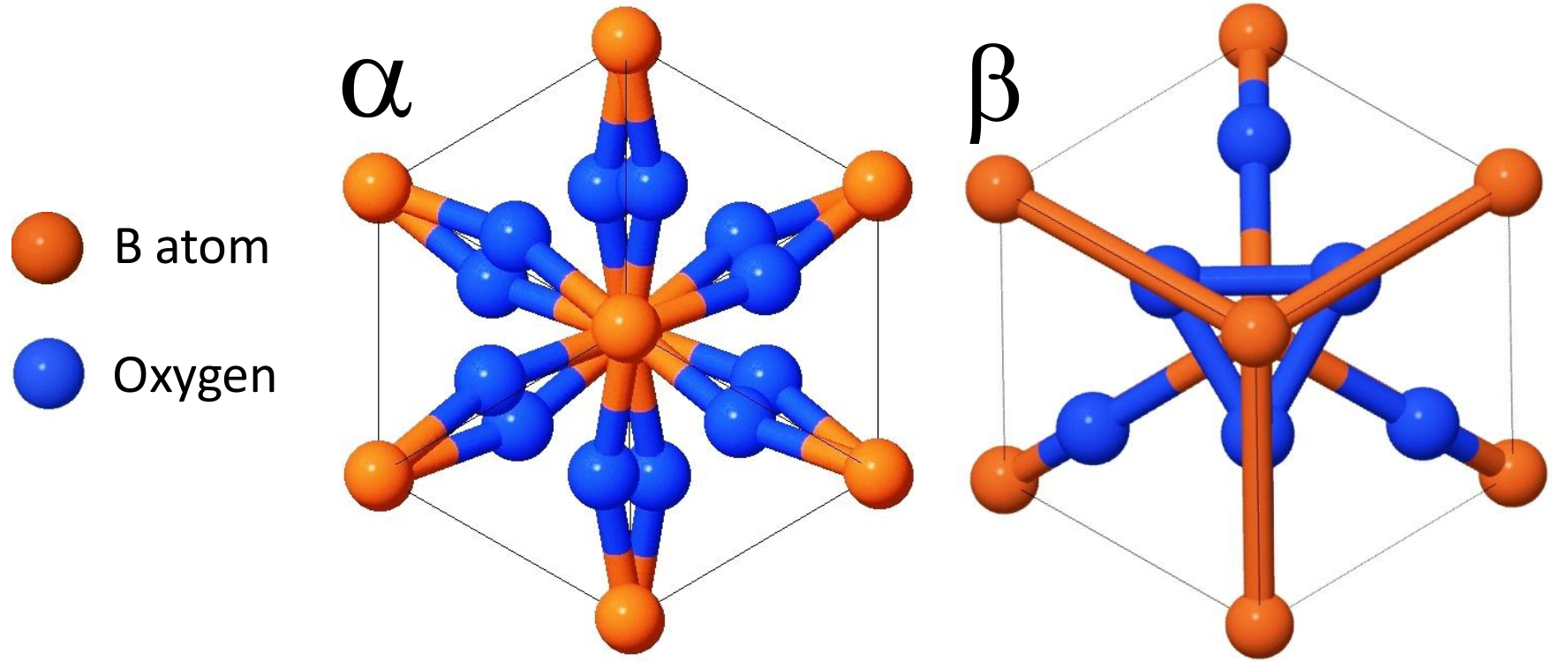}
\caption{Distortions $\alpha$ and $\beta$ 
  that reduce the ideal perovskite symmetry to R3c (along threefold axis).  The distortion $\beta$ combines with $\alpha$ at order $\alpha \beta$ to give both the observed counter-rotation and a linear magnetoelectric coupling.}
\label{fig} 
\end{figure}

Financial support from NSERC (R.d.S.) and WIN (J.E.M.) is acknowledged. \\

\noindent
Rogerio de Sousa$^1$ and Joel E. Moore$^2$\\
$^1$Department of Physics and Astronomy, University of Victoria, Victoria, BC V8W 3P6, Canada; 
$^2$Department of Physics, University of California, and Materials Sciences Division, LBNL,
Berkeley, CA 94720

\section{Explicit construction of $\alpha$ and $\beta$ distortions (Appendix to Comment on ``Ferroelectrically Induced Weak Ferromagnetism by Design'') }

It is convenient to use the following notation: Vectors written
without parenthesis denote components along the primitive vectors:
$x,y,z\equiv x \bm{a}+ y\bm{b} + z\bm{c}$. A convenient set of
primitive vectors for the 10-atom R3c unit cell is $\bm{a}=(0,1,1)$,
$\bm{b}=(1,0,1)$, and $\bm{c}=(1,1,0)$. Here vectors with parenthesis
are in a cartesian basis, and for convenience we choose the origin to
be on top of a B atom. Note that if $l,m,n$ is a set of integers, our
notation implies that $x+l, y+m, z+n$ is equivalent to $x,y,z$ apart
from a Bravais vector translation.

For a general ABO$_3$ perovskite with R3c symmetry, both A and B atoms
are located at Wyckoff 2a positions: The A atoms are located at
$x',x',x'$ and $x'+1/2,x'+1/2,x'+1/2$, while the B atoms are located
at $x'',x'',x''$ and $x''+1/2,x''+1/2,x''+1/2$. 

The six oxygens must be located at Wyckoff 6b positions: 
\begin{eqnarray}
x,y,z \;\; &z,x,y& \;\; y,z,x \nonumber\\
y+\frac{1}{2}, x+\frac{1}{2},z+\frac{1}{2} \;\;
&x+\frac{1}{2}, z+\frac{1}{2},y+\frac{1}{2}& \;\;
z+\frac{1}{2}, y+\frac{1}{2},x+\frac{1}{2} \;\;\nonumber
\end{eqnarray}

We may parametrize a set of \emph{paraelectric} R3c structures (under the definition of polarization assumed for this example in the text, in which the polarization is well described as translation of the anion and cation lattices; if the definition of ${\bf P}$ changes, then that of $\beta$ changes, but by counting such a $\beta$ always exists) with two degrees of freedom $\alpha$, $\beta$ according to:
\begin{eqnarray}
x'  &=& \frac{1}{4} \hspace{60pt} x'' =0 \nonumber\\
x   &=& -\frac{1}{4} + \alpha + \beta\hspace{20pt}
y   = \frac{3}{4}  - \alpha + \beta\hspace{20pt}
z   = \frac{1}{4}-2\beta\nonumber
\end{eqnarray}

This leads to the following atomic positions:
\begin{eqnarray}
\bm{R}_{\rm{A}} &=& \frac{1}{4}, \frac{1}{4}, \frac{1}{4}\hspace{20pt}
\frac{3}{4}, \frac{3}{4}, \frac{3}{4}\nonumber\\
\bm{R}_{\rm{B}} &=& 0,0,0 \hspace{26pt} 
\frac{1}{2}, \frac{1}{2}, \frac{1}{2}\nonumber\\
\bm{R}_{\rm{O}} &=& -\frac{1}{4} + \alpha +\beta , \frac{3}{4} -\alpha+\beta,
\frac{1}{4} -2\beta\hspace{10pt} \rm{and\;cyclic\;permutations}
\nonumber\\&&\hspace{8pt}\frac{5}{4}  
-\alpha +\beta, \frac{1}{4} + \alpha +\beta,
\frac{3}{4} -2\beta\hspace{10pt}\rm{and\;cyclic\; permutations}.\nonumber
\end{eqnarray}
Note that any equivalent lattice site could equally well be chosen for the oxygen locations, e.g., the choice above $-\frac{1}{4}, \frac{3}{4}, \frac{1}{4}$ is equivalent to $-\frac{1}{4}, -\frac{1}{4}, \frac{1}{4}$, or 
$\frac{3}{4}, \frac{3}{4}, \frac{1}{4}$, and so on.

The $\alpha$ and $\beta$ distortions defined in this way are
non-polar, i.e., orthogonal to ${\bf P}$. To see this, consider the electric dipole moment produced
by the oxygens,
\begin{equation}
P_{\rm{O}}=Q_{\rm{O}}\sum_{\rm{O}}
\bm{R}_{\rm{O}}=\sum_{l,m,n}[3+2(l+m+n)]\times 1,1,1.  
\nonumber
\end{equation}
This is seen to be independent of $\alpha, \beta$ and is equal to zero
when summed over a symmetric set of lattice vectors (the
electric dipole moment of the A and B atoms is also zero after a
suitable summation).  Direct computation shows that there is a counter-rotation by an angle proportional to $\alpha \beta$, as shown in Fig.~1 of the Comment.

The transformation properties under inversion on top of an A atom,
$I_A$, and inversion on top of a B atom, $I_B$, are straightforward to
compute from the atomic positions above. For $I_B$, just apply
inversion directly to the $\bm{R}_{\rm{O}}$ coordinates, while for
$I_A$ we must first subtract $1/4,1/4,1/4$, invert, and then add
$1/4,1/4,1/4$. The final result is simple to state: When $I_A$ is
applied, the original structure is recovered if we apply the
transformations $\alpha\rightarrow -\alpha$ and $\beta\rightarrow
-\beta$. Under $I_B$, the original lattice will be recovered only if
we apply $\alpha\rightarrow +\alpha$, and $\beta\rightarrow -\beta$.
In summary,
\begin{eqnarray}
I_A (\alpha, \beta) &=& (-\alpha,-\beta)\nonumber\\
I_B (\alpha, \beta) &=& (\alpha, -\beta). 
\end{eqnarray}


\begin{thebibliography}{99}

\bibitem{fennie} C. Fennie, \prl {\bf 100}, 167203 (2008).

\bibitem{edererfennie} C. Ederer and C. Fennie, J. Phys.: Condens. Matter {\bf 20}, 434219 (2008).



\bibitem{edererspaldin} C. Ederer and N. A. Spaldin, \prb {\bf 71}, 060401(R) (2005).

\bibitem{ravindran} P. Ravindran, R. Vidya, A. Kjekshus, H. Fjellvag, and O. Eriksson, \prb {\bf 74}, 224412 (2006).

\bibitem{kubel} F. Kubel and H. Schmid, Acta Cryst. {\bf B46}, 698 (1990).

\bibitem{note} The $c$ and $d$ $O-O$ bond lengths
 in \cite{kubel} indicate this counter-rotation of $1.4^{\circ}$.


\end{thebibliography}
\end{document}